\documentclass[preprint,showpacs,preprintnumbers,aps]{revtex4-1} 
\usepackage{graphicx}
\usepackage{dcolumn}
\usepackage{bm}%

\begin{document}

\preprint{arXiv: 0905.3605} 

\title{\bf General Relativistic effect on the energy deposition rate for
neutrino pair annihilation above the equatorial plane along the symmetry 
axis near a rotating neutron star}

\author{Ritam Mallick$^{w,}$\footnote{Email : ritam.mallick5@gmail.com},
Abhijit Bhattacharyya$^{z,}$\footnote{Email : abphy@caluniv.ac.in}, 
Sanjay K. Ghosh$^{x,y,}$\footnote{Email : sanjay@bosemain.boseinst.ac.in} 
 and
Sibaji Raha$^{x,y,}$\footnote{Email : sibaji@bosemain.boseinst.ac.in}}

\affiliation{$^w$Institute of Physics; Sachivalaya Marg;
 Bhubaneswar - 751005; Orissa; INDIA}
\affiliation{$^x$Department of Physics; Bose Institute;
 93/1, A.P.C Road; Kolkata - 700009; INDIA}
\affiliation{$^y$Centre for Astroparticle Physics and
Space Sciences; Bose Institute; 93/1, A.P.C Road; Kolkata - 700009;
INDIA}
\affiliation{$^z$Department of Physics; University of Calcutta;
 92, A.P.C Road; Kolkata - 700009; INDIA}

\date{\today}

\vskip 0.4in
\begin{abstract} 
The estimate of the energy deposition rate (EDR) for neutrino pair annihilation
has been carried out. The EDR for the neutrinos coming from the 
equatorial plane of a rotating neutron star is calculated along the rotation
axis using the Cook-Shapiro-Teukolsky (CST) metric. The neutrino trajectories and hence the neutrino
emitted from the disk is affected by the redshift 
due to disk rotation and gravitation. The EDR is very sensitive to the value of the temperature and its variation 
along the disk. The rotation of the star has a
negative effect on the EDR; it decreases with increase in rotational velocity.
\end{abstract}

\pacs{97.60.Jd, 26.60.+c}

\maketitle

\section{Introduction}

Neutron stars are objects formed 
in the aftermath of a supernovae. The central density of these stars can be as 
high as $10$ times that of normal nuclear matter. At such high density, 
any small perturbation,
{\it e.g.} spin down of the star, may trigger a phase transition from 
nuclear to quark matter
system. As a result, the neutron star may convert to a quark star or
a hybrid star with a quark core \cite{key-3a,key-3b}. It has been 
shown \cite{key-4} that such a 
phase transition \cite{key-5} produces a large amount of high energy neutrinos. 
These neutrinos (and antineutrinos) could annihilate and 
give rise to electron-positron pairs through the reaction $\nu {\bar \nu} 
\rightarrow e^+ e^-$. These $e^+ e^-$ pairs may further give rise to gamma 
rays which may provide a possible explanation of the observed GRB. 
Furthermore, the rotating neutron stars have been shown \cite{key-6} to produce 
the observed beaming effect of the GRB. At present, it is necessary to have 
a better 
understanding of the energy deposition in the neutrino annihilation to 
$e^{+} e^{-}$ process in the realistic neutron star environment.

Motivated by the delayed explosion of Type II supernovae, the EDR due to
the reaction $\nu {\bar \nu} 
\rightarrow e^+ e^-$ in the vicinity of a neutron star have 
been calculated \cite{key-7,key-8} based on Newtonian gravity, {\it i.e.}  
$( 2GM/c^2R ) << 1 $. Goodman {\it{et al.}} \cite{key-8} pointed out that 
the neutrino pair annihilation rate can be seriously
altered by the gravitational effects.
The effect of gravity was incorporated in 
refs. \cite{key-9,key-10}, but only for a static star. Slow rotation
was introduced by Prasanna and Goswami \cite{key-10a}.
GR effect on energy 
deposition rate for neutrino pair annihilation near a black hole was studied 
in ref. \cite{key-11,key-12} to explore the possible engine for GRBs.
There, the most probable candidate for the central engine was taken to be 
the accretion
disk around a black hole.
A detailed hydrodynamic simulation of the process was studied by 
Birkl {\it{et al.}} \cite{key-13}.

In an earlier paper \cite{key-14}, we have shown that the path of 
neutrino is important
for the study of EDR near a massive object. We have done a complete GR
calculation of the neutrino path for the most general metric describing a 
rotating star, and obtained its geodesic equation along the equatorial
and polar plane. The minimum photosphere radius (MPR) was
calculated for various stars along these 
two planes. This is the starting point of our present calculation. The MPR
serves as the lower limit from which we begin our calculation of the 
energy deposition rate (EDR) during the neutrino-antineutrino annihilation to 
electron-positron pairs.
Our aim in this work is to study, semianalyticaly, the relativistic 
effect on the EDR
above the equatorial plane for a star rotating along the polar axis. We
calculate the EDR along the rotation axis for the neutrinos coming from the 
equatorial plane of the star.

The relativistic effect consists of three factors: the gravitational redshift,
the bending of the neutrino trajectories and the redshift due to rotation.
The EDR is enhanced by the effect of neutrino bending;
however, the redshift due to disc rotation and gravitation reduces the EDR.
Thus these effects further complicate the estimation of EDR. In this work we
primarily focus on the effect of GR and rotation on the EDR.
The neutrinos released during phase transition will have very high energy
and will interact with the propagating medium. But first, let us calculate the
EDR only due to the thermal neutrinos, which have low energies and
therefore high mean free path. This EDR calculation will give us an
estimate of the EDR, and we will see how it compares with the energy
liberated during gamma ray bursts (GRB).
In order to discuss quantitatively the central engine of GRBs, we need a
comprehensive study of the formation, evolution, temperature dependence,
geometry of the star, the mechanism of energy deposition etc., most of which
are affected by the rotational and GR effect.

Our paper is arranged as follows. In the next section we discuss our 
equations of state (EOS) and 
describe the geometry of the star. In section III we formulate the algorithm
of the EDR calculation above the equatorial plane along the rotation axis.
Next we present our results and finally we discuss the results 
and summarize our work.

\section{Star structure}
The structure of the star is described by the CST metric given by \cite{key-15}
\begin{eqnarray}
ds^2 = -e^{\gamma+\rho}dt^2 + e^{2\alpha}(dr^2+r^2d\theta^2) 
+ e^{\gamma-\rho}r^2 sin^2\theta(d\phi-\omega dt)^2.
\end{eqnarray}
The four gravitational potentials, namely $\alpha, \gamma, \rho$ and 
$\omega$ are functions of $\theta$ and $r$ only. All the potentials 
have been solved for both static as well as rotating stars using the 
{\bf 'rns'} code \cite{key-16,key-17,key-18}.
For a rotating star, all the potentials become functions of both 
$r$ and $\theta$.

Tabulated equations of state (EOS) are needed to run the 
code. In this paper we have used hadronic EOS 
evaluated using the nonlinear Walecka
model \cite{key-19}. The Lagrangian in the model includes nucleons 
(neutrons and protons),
electrons, isoscalar scalar, isoscalar vector and isovector vector
mesons denoted by $\psi_{i}$, $\psi_{e}$, $\sigma$, $\omega^{\mu}$
and $\rho^{a,\mu}$, respectively. The Lagrangian also includes cubic
and quartic self interaction terms of the $\sigma$ field. The parameters
of the nonlinear Walecka model are meson-baryon coupling constants,
meson masses and the coefficient of the cubic and quartic self interaction
of the $\sigma$ mesons. The meson fields interact
with the baryons through linear coupling. The $\omega$ and $\rho$
meson masses have been chosen to be their physical masses. The rest
of the parameters, namely, nucleon-meson coupling constants 
and the coefficients 
of cubic and quartic terms of the $\sigma$ meson self interaction
are determined by fitting the nuclear matter
saturation properties, namely, the binding energy/nucleon (-16 MeV), 
baryon density
($\rho_{0}$=0.17 $fm^{-3}$), symmetry energy coefficient (32.5 MeV),
Landau mass (0.83 $m_{n}$) and nuclear matter incompressibility (300 MeV).

Using this tabulated EOS and a fixed central density, we run the 
{\bf{rns}} code to obtain all the gravitational potentials as a function of
$r$ and $\theta$, which thereby define the shape, mass, rotational velocity 
and other parameters of the star.
The shape of a fast rotating neutron star becomes oblate spheroid 
\cite{key-15}. The star gets compressed along the 
z-axis, on the other hand along x and y-axes it bulges by equal 
amounts, thereby making the polar radius smaller than equatorial radius.

\section{GR calculation}
In this section we study the GR effect on the neutrino pair annihilation rate.
The coordinate system is oriented such that 
the equatorial plane lies along
$\theta=\frac{\pi}{2}$ and the polar plane along $\theta=0$. We treat the 
equatorial plane as a disk
from which the neutrinos are coming out and depositing their energy along
different radial points on the rotation axis.  
Here we are interested in the EDR via neutrino pair 
annihilation near the rotation axis, $\theta =0$ \cite{key-11,key-12}. 
It has been shown earlier \cite{key-11} 
that in the absence of gravitation the $\theta$ dependence of the 
EDR is weak for small
values of $\theta$. Although the $\theta$ dependence of the gravitational 
effect may not necessarily be small, here we calculate the EDR at 
$\theta = 0$ and assume it to be approximately the same for small values of 
$\theta$.

The energy deposited per unit volume 
per unit time is given as \cite{key-7,key-8}

\begin{eqnarray}
\dot{q}(r,\theta)= \int \int f_{\nu}(p_{\nu},r) f_{\overline{\nu}}(p_{\overline{\nu}},r) (\sigma|v_{\nu}-v_{\overline{\nu}}|\epsilon_{\nu}
\epsilon_{\overline{\nu}}) 
\frac{\epsilon_{\nu} + \epsilon_{\overline{\nu}}}{\epsilon_{\nu} \epsilon_{\overline{\nu}}} d^3 p_{\nu} d^3 p_{\overline{\nu}},
\end{eqnarray}
where $f_{\nu}$ ($f_{\overline{\nu}}$) is the number density of neutrino 
(antineutrino), $v_{\nu}$ is the neutrino 
velocity, and $\sigma$ is the rest frame cross section for the process 
$\nu {\bar \nu} \rightarrow e^+ e^-$. 
The Lorentz invariant term is given by
\begin{eqnarray}
(\sigma|v_{\nu}-v_{\overline{\nu}}|\epsilon_{\nu} \epsilon_{\overline{\nu}}) 
= \frac{D{G_F}^2}{3\pi} (\epsilon_{\nu} \epsilon_{\overline{\nu}} - p_{\nu}.
p_{\overline{\nu}})^2
\end{eqnarray}
with 
\begin{eqnarray}
{G_F}^2 = 5.29 \times 10^{-44} cm^2 MeV^{-2}; \\
D = 1 \pm 4sin^2 \theta_w + 8 sin^4 \theta_w.
\end{eqnarray}
$sin^2 \theta_w = 0.23$. $(+)$ sign is for electron neutrino-antineutrino 
pair and $(-)$ sign for other pairs.

As the geometry of the equatorial plane is circular, we can decouple the
energy and angular dependence. Thus the rate of energy deposition is 
\begin{eqnarray}
\dot{q}(r)= \frac{DG_F^2}{3\pi }  F(r) \int \int f_{\nu}
f_{\overline{\nu}}(\epsilon_\nu + \epsilon_{\overline{\nu}}) 
\epsilon_\nu^3 \epsilon_{\overline{\nu}}^3 d\epsilon_{\nu} 
d\epsilon_{\overline{\nu}} 
\end{eqnarray}

where, $F(r) $, the angular integral, is given by \cite{key-12}
\begin{eqnarray}
F(r)=\frac{2\pi^2}{T_{eff}^9} (\frac{e^{\gamma + \rho}}{e^{\gamma + \rho} -
r^2 \omega^2 sin^2\theta e^{\gamma - \rho}})^4 (2 {\int_{\theta_m}^{\theta_M}}
d\theta_{\nu} {T_0}^5 sin\theta_{\nu}  {\int_{\theta_m}^{\theta_M}}
d\theta_{{\bar \nu}} {T_0}^4 sin\theta_{{\bar \nu}} \nonumber \\
+ {\int_{\theta_m}^{\theta_M}}
d\theta_{\nu} {T_0}^5 sin ^3 \theta_{\nu}  {\int_{\theta_m}^{\theta_M}}
d\theta_{{\bar \nu}} {T_0}^4 sin^3 \theta_{{\bar \nu}} + 2 {\int_{\theta_m}
^{\theta_M}}
d\theta_{\nu} {T_0}^5 cos^2 \theta_{\nu} sin\theta_{\nu}  {\int_{\theta_m}^
{\theta_M}} d\theta_{{\bar \nu}} {T_0}^4 cos^2 \theta_{{\bar \nu}} sin\theta_
{{\bar \nu}} \nonumber \\
-4 {\int_{\theta_m}^{\theta_M}}
d\theta_{\nu} {T_0}^5 cos \theta_{\nu} sin\theta_{\nu}  {\int_{\theta_m}^
{\theta_M}} d\theta_{{\bar \nu}} {T_0}^4 cos \theta_{{\bar \nu}} sin\theta_
{{\bar \nu}}) 
\label{fr}
\end{eqnarray}
where $\theta_{\nu}, \theta_{\overline{\nu}}, \theta_m$ and $\theta_M$ are 
explained later.
 
We calculate the energy integral numerically, integrating it from $0$ to 
$\infty$. Both the angular and energy integrals have temperature 
dependence which 
is affected by the GR and rotational effects.
$T_{eff}$ is the effective temperature of the disc that is observed in the 
comoving frame. The value of $T_0$ is the temperature of the disc that is 
observed at infinity. 
In the locally nonrotating frame a neutrino moves normally to the direction of
the disk motion. In that case, the temperature suffers redshift both due to 
disk rotation and gravitation. Therefore the
two temperature $T_0$ and $T_{eff}$ are related
\cite{key-12}, as
\begin{eqnarray}
T_0=\frac{T_{eff}}{\Gamma} \sqrt{g_{tt}-\frac{{g^2}_{t\phi}}{g_{\phi \phi}}}.
\end{eqnarray}
$g_{tt},g_{t\phi}$ and $g_{\phi \phi}$ are calculated from the 
metric. $\Gamma$ is the Lorentz
factor defined as $\Gamma = \frac{1}{\sqrt{1-v^2}}$ with $v$ being
$v=(\Delta-\omega)r sin\theta e^{-\rho}$, $\Delta$ the rotational 
velocity of the star.

In an earlier paper \cite{key-14} a detailed GR 
calculation of the neutrino
path has been done.  For the present coordinate system, the 4- momentum 
\cite{key-12} is
given by the equation,
\begin{eqnarray}
g_{\mu\nu}p^{\mu}p^{\nu} + {\mu_m}^2 = 0,
\end{eqnarray}
where $\mu_m$ is the rest mass of the particle and $p^{\mu}=\frac{dx^{\mu}}
{d\lambda}$, $\lambda$ being the affine parameter. After a bit of algebra the
final geodesic equation can be written as,
\begin{eqnarray}
e^{2\alpha}[\frac{A(r,\theta)}{A(r,\theta)+\omega}]^2
\frac{e^{(\gamma - \rho)} r^2sin^2 \theta}{e^{2\alpha}} tan^2 \theta_r
[\omega (1-\omega b)+\frac{e^{2\rho}b} {r^2sin^2\theta}]^2 \nonumber \\
- e^{(\gamma + \rho)}(1-\omega b)^2 + \frac{b^2}{r^2sin^2\theta }
e^{(\gamma + 3\rho)} = 0.
\end{eqnarray}
$A(r,\theta)$ is defined as
\begin{eqnarray}
A(r,\theta)=\frac{L}{E-\omega L}\frac{e^{2\rho}}{r^2sin^2\theta} \nonumber
\end{eqnarray}
where $E$ and $L$ are the total energy and total angular momentum of the 
neutrino measured from infinity.

This equation can be solved using the potentials obtained from
{\bf rns} code to obtain a minimum
radius $r=R_{MPR}$, the minimum photosphere radius, below which a
massless particle (neutrino) emitted
tangentially to the stellar surface ($\theta_R =0$) would be
gravitationally bound. This is also the minimum radius from which the 
neutrinos can come out from the disk.

From an angle $\theta_{\nu}$ at a point $(r,0)$ (refer fig. 1), we can trace 
back a trajectory of the neutrino to its emission point on the disc. Thus the
temperature $T_0$ depends on $\theta_{\nu}$, and therefore $T_{eff}$ appears
in the integration of $\theta_{\nu}$ in the angular integral. 
The neutrinos are coming out of the disk and depositing their energy along the 
rotation (symmetry) axis of the disk. The minimum radius from which 
the neutrinos can come out is given by MPR. The outermost point from which 
the neutrinos come
out is the surface of the disc ({\it{i. e.}} the surface 
of the star along the 
equatorial plane). Figure 1. shows the layout of our problem, where neutrinos 
coming out from the disk are depositing their energy along the 
rotation axis. Neutrinos emitted from the disk at 
$(r,\theta)=(R,\frac{\pi}{2})$, arrive at point $(r,0)$ at the rotation axis.
A neutrino coming from the MPR ($R_{MPR}$) subtend angle $\theta_m$ 
with the rotation axis, the 
minimum value of $\theta_{\nu}$ and that coming from the outer surface of 
the disk ($R_{Sur}$) subtends angle $\theta_M$, the maximum value of 
$\theta_{\nu}$. The angle $\theta_{\nu}$ is defined as the angle between 
position vector $r$ and $p_{\nu}$. The minimum and the
maximum angles are given by
\begin{eqnarray}
sin\theta_m=\frac{(R_{MPR}e^{-\rho_{MPR}})(1+\omega r e^{-\rho})}
{(1+\omega_{MPR}R_{MPR}e^{-\rho_{MPR}})(re^{-\rho})}
\end{eqnarray}
and
\begin{eqnarray}
sin\theta_M=\frac{(R_{Sur}e^{-\rho_{Sur}})(1+\omega r e^{-\rho})}
{(1+\omega_{Sur}R_{Sur}e^{-\rho_{Sur}})(re^{-\rho})}.
\end{eqnarray}

The EDR is now calculated by integrating $\dot{q}$ over the proper volume.
We calculate the amount of energy deposited along the rotation axis 
within an angular opening of about ten degrees \cite{key-11}.
The point on the rotation axis from 
where we start our calculation is $r= R_{MPR}$.
Therefore the EDR is given by
\begin{eqnarray}
EDR=2\pi {\int_{r_{n}}^{r_{n+1}}}dr {\int_{\theta_1}^{\theta_2}} \dot{q}(r) 
r^2sin\theta \frac{e^{(2\alpha + (\gamma-\rho)/2)}}{\sqrt{1-v^2}} d\theta.
\end{eqnarray} 
The integration is done over radial bin of $100m$, {\bf{i.e.}} 
$r_{n+1}-r_n=100m$, and an angular width of ten degree, {\bf{i.e.}} 
$\theta_1=0^{\circ}$ and $\theta_2=10^{\circ}$.

The observed luminosity at infinity $L_{\infty}$ of the neutrino pairs
annihilation is given by
\begin{equation}
L_{\infty}=[\frac{{g^2}_{t\phi}(R)}{g_{\phi \phi}(R)}-g_{tt}(R)]L(R)
\end{equation}
\label{lumin_surface}
The neutrino luminosity at the MPR is
\begin{equation}
L(R)=L_{\nu}+L_{\bar{\nu}}=\frac{7}{16}4 \pi R^2 a {T_{eff}}^4(R)
\end{equation}
\label{lumin_mpr}
where $a$ is the radiation constant.

In general the total neutrino luminosity will depend on the neutrino emissivity and can be evaluated with
the volume integral of angle averaged neutrino emissivity \cite{berdermann}. Since some of the neutrinos will be depositing
their energy in $e^{+} e^{-}$ pairs, the total emissivity and hence neutrino luminosity will decrease accordingly. 
One should also consider the actual number of neutrinos trapped, the number of which will depend on the neutrino energy, 
matter density and temperature. In eqn.[14] we have considered only the total luminosity as observed due to the red-shift 
in temperature.
Moreover, eqn. [15] is valid for the case where neutrino chemical potential inside MPR is
small enough compared to the temperature. Hence, the above values of luminosity, given by eqns. (14) and (15) 
will provide with a lower limit. 

We consider a NS which is of the order of a minute old. The average
neutrino energy is of the order of few $MeV$, and the neutrino mean free 
path is of the order of the radius of the star \cite{key-prakash}.
We start our calculation with three different central temperature, $1 MeV$, 
$5 MeV$ and $10 MeV$. We consider the neutrinos comes out from the disk 
beyond the MPR (which 
is $4 Km$). At this distance the star is much less dense than the central 
part, actually if the central density is $7$ times the nuclear saturation 
density, the density at MPR is less than $4$ times nuclear matter 
saturation density.
In our problem we are considering only 
the thermal neutrinos. Following the prescription given in ref.
\cite{iwamoto}, the mean free path of the neutrinos with such energy and  
the given density is of the  
order of kilometers. Other calculations \cite{fabri,giai} do 
not alter the result much and the mean free path of the neutrinos 
for density and temperature of the star considered in the present study 
remains of the order of kilometers. Therefore the 
neutrinos can travel quite a distance within the star and deposit their 
energy along the rotation axis. 

\section{Results}
We choose the central density of the star to be $1 \times 10^{15} gm/cm^3$,
for which the Keplerian velocity is $0.61 \times 10^4 s^{-1}$. For comparison 
we have also done the calculation for a slow star 
($0.4 \times 10^4 s^{-1}$). For the Keplerian star the equatorial radius
is $16 km$, which serves as the outermost point of the disk. The innermost 
point from which the neutrinos can contribute in the EDR is the MPR, which
is $4 km$ from the center. The same two points for the slowly rotating star
are $12 km$ and $4.7 km$ respectively \cite{key-14}.
 
Before discussing our results, we would like to mention that our present approach is a 
simplification of the actual scenario and the effect of neutrino inside NS may 
be more complicated. The trapping of neutrino inside the MPR may result 
in a change in the temperature profile as well as structure of the NS. 
For example, during the Neutrino cooling period, the external layer near the MPR, where 
the neutrino trapping reaches highest efficiency, temperature may become higher than 
the temperature in the inner layers. This effect will then lead to an inflow of heat 
to the interior through some processes \cite{class}. This could then influence the internal structure of the 
NS.

It has also been found \cite{vidana} that the presence of neutrino changes the composition of  
matter significantly with respect to the neutrino free case; matter becomes more proton reach and 
hyperons appear at higher densities. Since the presence of hyperons softens the EOS, in presence 
of neutrino trapping EOS will become stiffer. 

In general the neutrino mean free path decreases with temperature (at fixed density) and 
density (for fixed temperature)\cite{key-prakash,reddy}. So the high neutrino trapping and hence the 
increase in the temperature near the MPR will result in the decrease in the mean free path 
thereby changing the temperature profile of the star \cite{berdermann}. 

The emissivity (energy emitted per unit volume per unit time)may be taken to be proportional to
$T^6$ and ${n_B}^{1/3}$ \cite{lattimer}, $n_B$ being the baryon density. This is the amount of 
energy (per unit volume per unit time) deposited in the matter, if all the neutrinos are trapped. 
For a nonrotating spherical NS, made up of nucleons only, the density $n_B$ may be taken to be proportional to $r^{-3}$, 
$r$ being the radius of the star. Hence the emissivity varies as $1/r$. If we assume all the energy 
deposited goes into thermal energy, then temperature due to this deposition will also vary as 
$1/r$. So there will be a temperature gradient (TG) of $1/r$ along the disk in the present problem. 
Since a rotating star changes its shape and becomes an oblate spheroid, the density and hence the 
temperature variation may change accordingly.

Here we first assume an isothermal disk, {\it{i.e.}} there is
no TG along the disk. 
Figure 2 shows the variation of the EDR along the rotation
axis of the star for three different temperatures ($1MeV$, $5MeV$ and $10 MeV$).
The neutrino mean energy is in the range between $1-10MeV$.
Initially the EDR increases with distance and it reaches a maximum value at 
a distance of $16 km$ and after that it falls off gradually. The EDR is 
maximum for central temperature
$T_C = 10 MeV$ and minimum for $T_C = 1 MeV$, which shows that
the EDR increases with increase in temperature of the disk. For $T_C = 10 MeV$ 
the maximum EDR is at a point about $16 km$ and the value of EDR is  
$3.3 \times 10^{47} erg/s$, and the total energy deposited per second 
in the vicinity of the star (calculated by integrating EDR) is of the order of 
$10^{49} erg/s$. If we further increase the temperature to $20 MeV$ the total 
energy deposited per second near the star is of the order of 
$10^{51}-10^{52} erg/s$, 
close to that of the energy liberated during GRBs. At this temperature
the observed luminosity $L_{\infty}$ is of the order of the EDR. This 
implies that the neutrino becomes optically thick for the pairs annihilation.
For $T_C=10 MeV$ the EDR is $0.1$ percent of the neutrino luminosity 
($L_{\infty}$ is $9\times 10^{51} erg/s$).

Let us try to understand the peculiar nature of initial increase in EDR
with distance. From figure 3, for the isothermal disk
the temperature $T_0$, seen by an observer at infinity, 
increases as the distance from the star increases. 
In this case
$T_0$ becomes hotter with distance and therefore it approaches $T_{eff}$ 
(equal to $T_C$ for isothermal disk) at 
large distance. 
As the disk is isothermal everywhere along the disk its temperature is 
constant. At close distance from the center of the disk along the 
rotation axis the contribution from the the inner part of the disk is greater
and at larger distance the contribution is mostly from the outer part of 
the disk. This does not make any difference for the isothermal disk as all
along the disk the temperature is constant. 
As we go to a larger distance the GR effect reduces and $T_0$ 
approaches $T_{eff}$.
But for the disk with TG, $T_0$ will decrease with distance, unlike the above
isothermal case. This is because as we go to a larger distance along the
rotation axis the contribution from the outer part of the disk will dominate.
For the disk with TG as we move outward along the disk 
its temperature decreases 
rapidly and is much smaller than the central temperature. Although at a large 
distance GR effect will become less, $T_0$ will still decrease as 
here the temperature contribution comes mainly from the outer 
surface of the disk which is at a much lower temperature.

Looking at figure 4, for the isothermal disk, we see that $F(r)$ 
(eqn. \ref{fr}) decreases with distance and falls to zero after $25 km$. 
For the disk with temperature gradient we see that $F(r)$ 
initially increases and after
about $8 km$ it starts to fall, unlike the isothermal case where it
decreases throughout.

For an isothermal disk, initially the 
temperature dominates and EDR increases but after  
some distance (where $F(r)$ decreases but temperature increases) $F(r)$ 
becomes the dominating factor and the EDR decreases. The disk temperature
observed at infinity is higher at outer region and the EDR increases
with temperature. On the 
other hand the Doppler effect due to disk rotation reduces the EDR.
The EDR calculated non relativistically at the MPR is half that 
of the relativistic case \cite{key-7,key-8}.

The EDR via neutrino pair annihilation is strongly dependent on disk
temperature. Therefore let us calculate the pair annihilation of neutrinos
emitted from the disk with TG. As discussed previously one of the other 
extreme case will be the disk with $TG=1/r_d$ ($r_d$ is the radial distance
along the disk). For
comparison we also plot curves with $TG = 1/{r_d}^{1/2}$.
Due to variation of disk
temperature, the temperature at the MPR (initial temperature from which
our calculation starts) is much less than that of the central temperature.
For $T_C=10 MeV$ and $TG=1/r_d$ the $T_0$ at MPR is less than
$2 MeV$ and the temperature falls very rapidly and reaches
negligible values beyond $50 km$. The case of
$TG=1/{r_d}^{1/2}$ the situation is not so pessimistic. For the hotter 
star ($T_C=10MeV$), $T_0$ at the
MPR is above $3 MeV$ and does not fall to zero at large distances.

Figure 5 shows the variation of EDR, for disk with TG, 
along the rotation axis of the star for
two different central temperatures $5 MeV$ and $10 MeV$.
For $T_C=10 MeV$ with $TG = 1/r_d$ (surface temperature is $0.4 MeV$) , 
the initial value of EDR is
$10^{40} erg/s$ and it falls very sharply
to zero at a distance of $50 km$. The total EDR is near to $10^{41} erg/s$.
The disadvantage of the TG disk is that the EDR is $10$ percent of the 
neutrino luminosity ($L_{\infty}=9\times 10^{41} erg/s$), thereby again 
imposing the problem of neutrino opacity and also the baryon contamination 
problem \cite{key-12}.
For $T_C=5 MeV$ the fall is even sharper. For
$TG=1/{r_d}^{1/2}$ the fall in EDR is much slower and near the MPR
the EDR, for
central temperature $T_C=10 MeV$ (surface temperature is $2 MeV$), 
is of the order of
$10^{43} erg/s$. For the colder disk the initial EDR is much smaller.

These curves show that the EDR is very sensitive to temperature
as well as to TG.
At smaller $r$, temperature is the governing factor in
determining the nature of EDR, therefore EDR decreases. 
As $r$ increases, $F(r)$, which strongly decreases with $r$,
becomes the dominant factor and EDR decreases further.
Hence, here the EDR falls off much faster than the
isothermal case. 

The value of EDR increases if we go inwards from the MPR.  
If we go close towards the center, the EDR increases due to 
the increase in temperature of the disk. Near the center the value of EDR for
non relativistic calculation and that with GR effect along with TG becomes 
the same. But as we go outwards the value of the EDR for the GR calculation
becomes less than the non-relativistic case. This is because, for 
GR calculation, the disk temperature falls as we go outwards along the disk.

The slowly rotating star has a much smaller disk
(smaller equatorial plane) and its MPR is $4.7 km$, larger than the keplerian
star. The gross nature of the EDR 
is more or less same, but the value of maximum EDR is 
slightly larger.
The temperature and $F(r)$ plot is about the same, only the maximum value of 
$F(r)$ being slightly larger.
Due to this reason the EDR for the slow star is greater (twice)
than that for the keplerian star. Therefore the rotational velocity seems to have 
some negative effect on the EDR and as the rotational velocity increases the EDR 
decreases.
The other important conclusion from the above discussion is that the 
isothermal disk is more advantageous while considering the opacity and
baryon contamination problem (as for isothermal disk the ratio of EDR on
$L_{\infty}$ is minimum).

\section{Summary and Discussion}
In this article we have investigated the GR effect on the EDR 
of $\nu {\bar \nu} \rightarrow e^+ e^-$ reaction in a rotating 
neutron star described by  CST metric.
Here we have calculated the EDR for the neutrinos coming out of the 
equatorial disk and depositing energy along the rotation axis of the star, 
above the equatorial plane. 
The bending of the
neutrino trajectories, and the redshift due to disk rotation 
along with gravitation 
has been taken into consideration. 

We find that for an isothermal disk, initially the EDR increases with
distance and reaches a maximum value and then decreases with distance. 
This is due to the fact that there is 
a competition between the temperature and the Doppler effect due to disk 
rotation (characterized by $F(r)$). The 
temperature observed by the observer at infinity ($T_0$) increases and at 
large distance becomes same as that of the effective temperature ($T_{eff}$) 
whereas $F(r)$ 
decreases with increase in distance from the disk. In the case of disk with 
TG, the EDR falls off very quickly with distance.
For $TG=1/r_d$ the slope with which EDR falls is much greater than that for
$TG=1/{r_d}^{1/2}$.
The $F(r)$ increases initially but it also decreases after some distance 
and falls very near to zero at about $50 km$. Both the isothermal and TG 
picture shows that
the EDR is a very sensitive function of temperature, and as temperature 
decreases the EDR also decreases. 
We also conclude that the isothermal disk is more advantageous in avoiding
opacity and baryon contamination problem.
The slow star shows more or less the
same nature except for the fact that EDR for the slow star is about twice 
that for the keplerian star. The rotational effect reduces the EDR.

The method in which the angular integral is calculated is somewhat similar
to that mentioned in ref. \cite{key-11,key-12}. In that paper, 
the authors had calculated
the energy deposition due to neutrino-antineutrino annihilation above a
black hole. The Kerr metric describes the black hole, which is flat disk,
and the neutrinos emitted from the disk contribute to the deposited energy.
The calculation of the energy integral is different in our case and the
star is at much lower temperature. In our discussion the neutrino are emitted
from the equatorial plane of the star, which is a circular disk, and deposit
their energy along the rotation axis of the star (described in fig. 1). The
innermost point from which the neutrino contribute is the MPR and the
outermost point is the star surface. We have calculated the EDR 
for the neutrinos coming out from the equatorial plane. 
If we consider the star to be spherical, the total 
EDR for the entire star can be obtained multiplying the EDR calculated for 
the equatorial plane by $2\pi$, for which the EDR comes out to 
be of the order of 
$10^{50}ergs/s$. If we consider long duration GRB this EDR is quite close 
to that of the energy liberated 
during the GRB ($\sim10^{52}ergs$). 

To summarize, we find that the EDR is very sensitive both 
to the temperature and TG of the disk. The deposition energy is contributed
mainly by the neutrinos arriving from the central region where the 
temperature is higher. The maximum energy is deposited near the surface. 
As we move outwards the deposition energy reduces and
is smaller than that of the non relativistic calculation. The maximum 
amount of energy deposited for an isothermal disk is few times 
$10^{49} ergs/s$, within a small angle. The isothermal disk is the
most advantageous while avoiding opacity and baryon contamination problem.
Taking the disk to be isothermal a rough estimate of a spherical 
star provides EDR quite close to that of the
energy liberated during a long duration GRB.
The total energy deposition from a rotating neutron star would also depend on 
the off-axis contributions, but expected that most of the contribution will
be along the rotation axis (if it coincides with the magnetic poles), as the
particles comes out of the star along the magnetic poles. 
The high energy neutrinos created during phase transition may
further change the EDR.
To get an estimate of the total energy deposition, 
one needs to do a detailed
numerical simulation taking into account all possible contributions. 
Such a calculation is in progress and we hope to report it in the near future.

\acknowledgments{ A.B thanks UGC for financial support. S.K.G. and S.R thank DST, 
Govt of India, for financial support. It is also partly supported by grant no.
SR/S2HEP12/2007, funded by DST, India.}

\clearpage 

\begin{figure}
\vskip 0.2in
\centering
\includegraphics[width=5 in]{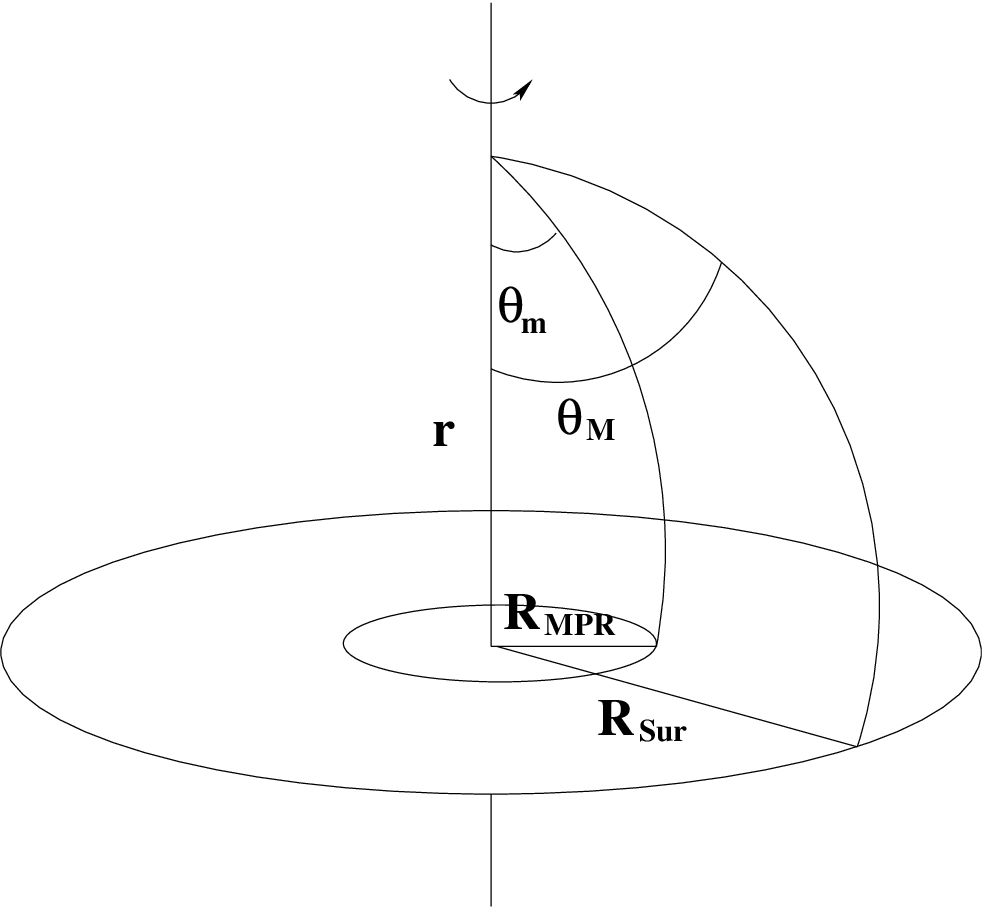}
\caption{Figure showing the layout of our problem. The neutrinos
(antineutrinos) coming out of the equatorial plane, at an angle ($\theta_M$ -
$\theta_m$) and depositing their energy along the rotation axis. }
\end{figure}

\begin{figure}
\centering
\includegraphics{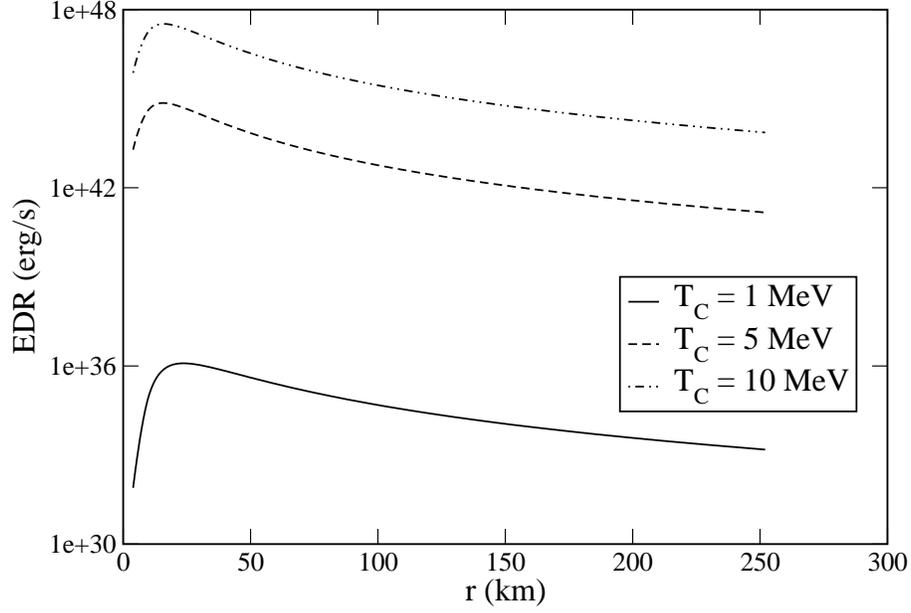}
\caption{Variation of EDR with distance along the rotation axis for the 
keplerian star having isothermal disk with three different central temperature.}
\end{figure}

\begin{figure}
\vskip 0.2in
   \centering
\includegraphics{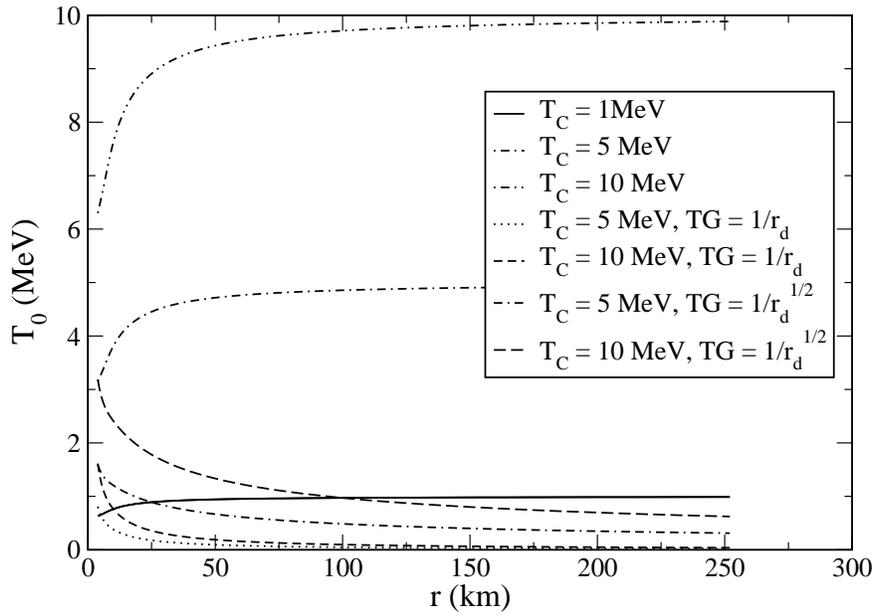}
\caption{Variation of temperature $T_0$ with distance along the rotation axis 
of the keplerian star. $T_0$ is plotted for both isothermal disk and for disk
having two different temperature gradient.}
\end{figure}

\begin{figure}
\centering
\includegraphics{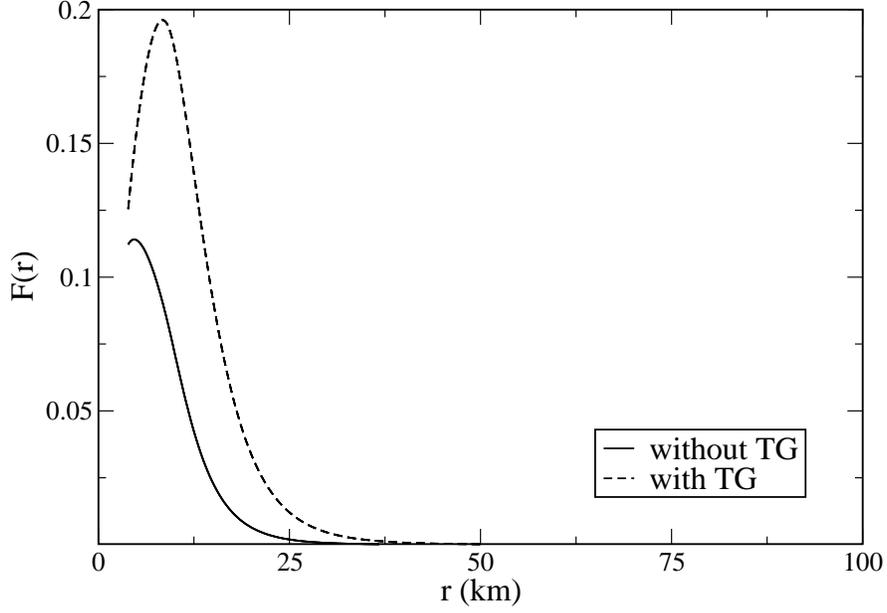}
\caption{Variation of $F(r)$ with distance along the rotation axis of the 
keplerian star. $F(r)$ is plotted both for isothermal disk and disk with
temperature gradient.}
\end{figure}

\begin{figure}
\vskip 0.2in
\centering
\includegraphics{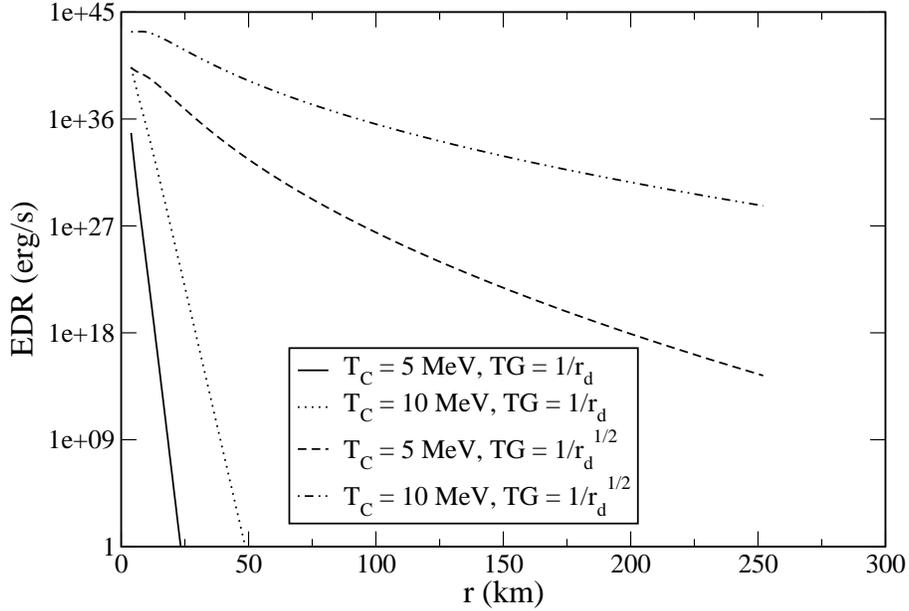}
\caption{Variation of EDR with distance along the rotation axis for the 
keplerian star. The plot have been done with disk having two different TG 
and the curves are plotted with two different central temperature.}
\end{figure}

\end{document}